\documentclass[12pt,a4paper]{article}
\usepackage[utf8]{inputenc}
\usepackage[T1]{fontenc}
\usepackage{geometry}
\usepackage{array}
\usepackage{multicol}
\usepackage{comment}
\usepackage{amsopn,amsmath,amssymb,amsthm,amstext,amsfonts,algorithmic,psfrag}
\usepackage{amssymb}
\usepackage{enumitem}
\usepackage{mathtools}
\usepackage[english]{babel}
\usepackage{hyperref}
\hypersetup{colorlinks=true, citecolor=blue, linkcolor=blue}
\usepackage[round]{natbib}
\setcitestyle{aysep={,}} 
\usepackage{authblk}
\usepackage{graphicx}
\usepackage{mwe}    
\usepackage{subfigure} 
\usepackage{float}
\usepackage{caption}
\usepackage{algorithm}
\usepackage{algorithmic,latexsym}
\usepackage{placeins,enumitem}
\renewcommand{\thetable}{\arabic{table}}

\usepackage{eso-pic}
\newcommand\AtPageUpperMyright[1]{\AtPageUpperLeft{%
 \put(\LenToUnit{0.5\paperwidth},\LenToUnit{-1cm}){%
     \parbox{0.5\textwidth}{\raggedleft\fontsize{9}{11}\selectfont #1}}%
 }}%
\newcommand{\conf}[1]{%
\AddToShipoutPictureBG*{%
\AtPageUpperMyright{#1}
}
}

\title{\LARGE \bf
Development of Decision Support System for Effective COVID-19 Management}

\author[1]{Shuvrangshu Jana}
\author[2]{Rudrashis Majumder}
\author[3]{Aashay Bhise}
\author[4]{Nobin Paul}
\author[5]{Stuti Garg}
\author[6]{Debasish Ghose}

\affil[1]{Post-doctoral Fellow, Department of Aerospace Engineering, Indian Institute of Science, Bangalore,
		{\tt\small shuvrangshuj@iisc.ac.in}}

\affil[2]{Ph.D. Student, Department of Aerospace Engineering, Indian Institute of Science, Bangalore, 
		{\tt\small rudrashism@iisc.ac.in}}
	
\affil[3]{Project Associate, Department of Aerospace Engineering, Indian Institute of Science, Bangalore, 
		{\tt\small meetaashay3@gmail.com }}

\affil[4]{Ph.D. Student, Department of Aerospace Engineering, Indian Institute of Science, Bangalore, 
		{\tt\small nobinpaul@iisc.ac.in}}

\affil[5]{Project Assistant, Department of Aerospace Engineering, Indian Institute of Science, Bangalore, 
		{\tt\small stutigarg9@gmail.com }}

\affil[6]{Professor, Department of Aerospace Engineering, Indian Institute of Science, Bangalore, 
		{\tt\small dghose@iisc.ac.in}}		
		
\date{}                     
\begin{document}
\maketitle 
\conf{$5^{th}$ World Congress on  Disaster Management\\ IIT Delhi, New Delhi, India, 24-27 November 2021}


\begin{abstract}
This paper discusses a Decision Support System (DSS) for cases prediction, allocation of resources, and lockdown management for managing  COVID-19 at different levels of a government authority.  Algorithms incorporated in the DSS are based on a data-driven modeling approach and independent of physical parameters of the region, and hence the proposed DSS is applicable to any area. Based on predicted active cases, the demand of lower-level units and total availability,  allocation, and lockdown decision is made. A MATLAB-based GUI is developed based on the proposed DSS  and could be implemented by the local authority. 
\end{abstract}


\section{Introduction}
 Management of COVID-19 involves optimal lockdown planning, estimation and arrangement of critical medical resources, and allocation of resources among the units in an optimal manner.  The human intervention for this set of activities is sometimes not optimal because of bias and other inaccuracies. Hence, building an autonomous decision support system (DSS) to handle all the activities related to disaster management can address the problems arising from the situation more effectively.

In the past, developing DSS has been given much importance by disaster researchers. Some early works of literature mainly focus on designing the decision support system for the influenza pandemic response. \cite{jenvald2007simulation} propose a simulation framework to simulate the pandemic environment and relate it to the decision support system (DSS) for influenza preparedness and response. \citep{fair2007integrated} developed another simulation setup to model the time-dependent evolution of influenza.\citep{arora2012decision} focus on the complexities associated with the decision-making during the outbreak of influenza pandemic. \citep{araz2013simulation} design a DSS to take decisions on the closure and reopening of schools to minimize influenza infection. \citep{araz2013integrating} present a general multi-criteria decision-making framework to make preparedness plans with DSS in order to integrate the estimation of important epidemiological parameters. \citep{fogli2013knowledge} develops a decision system for emergency response during a pandemic. \citep{shearer2020infectious} incorporates dynamic information about the pandemic from the situational awareness framework and integrates it into the broader DSS for pandemic response. Authors in \citep{phillips2020supporting} mainly focus on decision-making during the stressful outbreak of COVID-19.  \citep{currion2007open} talk about the significance of an open-source software model for disaster management. A brief introduction to Sahana software is also given in this paper. \citep{iyer2006important} discuss the important elements of disaster management and the development of a software tool to facilitate disaster DSS. Several other papers like \citep{li2013case, shukla2012considering} also focus on the software modeling for emergency and disaster management.

Researchers of different fields focused on developing DSS to mitigate the effect of the COVID-19 pandemic outbreak in 2020. \citep{guler2020decision} find the solution for the problem of shift scheduling of the physicians by using mixed-integer programming and then using it in a DSS. In \citep{sharma2020multi}, a multi-agent intelligent system is developed for decision-making to assist the patients. \citep{hashemkhani2020application} address hospital location selection problem for COVID-19 patients using gray-based DSS. The paper \citep{govindan2020decision} develops a practical decision support system depending on the health practitioners' knowledge and fuzzy information system to break the chain of COVID infection. \citep{marques2021prediction} uses AI-based prediction model for decision making in COVID-19 situation.


The DSS related to COVID-19 reported in the literature is mostly focused on specific activities related to resource management by medical personnel in a  hospital environment.  However, an integrated DSS  covering estimation, allocation, and lockdown management for government authority will help efficient disaster management. This paper describes autonomous  DSS that addresses prediction, allocation, and optimal lockdown management for efficient management of COVID-19 in India. The algorithms incorporated in DSS are scalable and flexible, and thus it applies to any level of a government authority. The algorithms are based on a short-term prediction of COVID-19 cases using the time series data of reported cases. A graphical GUI is developed for decision making and the GUI inputs are demand/availability of critical items and total cases, recovered cases, and deceased cases. The proposed DSS could help the authorities to take crucial allocation and lockdown decisions in an optimal manner. 
 
 The rest of the paper is described as follows: Section \ref{DSS} describes the current status of the decision-making process in COVID-19 and the requirement of DSS.  The overview of the algorithms incorporated in DSS is presented in  Section \ref{algorithm}. The description of the GUI of DSS is shown in Section \ref{GUI}.

\section{ COVID-19 management}  \label{DSS}
\begin{figure*}[htb!]
      \centering
      \includegraphics[width=\linewidth, keepaspectratio]{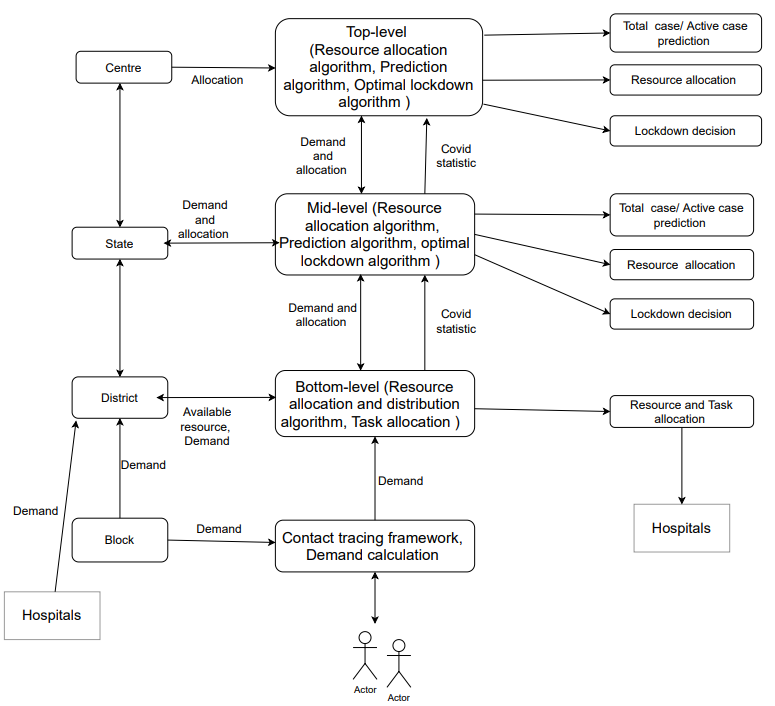}
      \caption{COVID management at different level}
      \label{fig:DSS_covid_software.png}
  \end{figure*}

The important tasks of a COVID management authority are the prediction of COVID-19, computation of demand, allocation and distribution of critical items, and decision making for the level of lockdown. In general, COVID-19 is managed through different hierarchical levels of authorities, and each level has different responsibilities.  COVID-19 management structure in the context of India is shown in  Fig. \ref{fig:DSS_covid_software.png}. In this case, COVID  management is performed through centre, state, district, and block hierarchy. Different responsibilities at each level are also mentioned. At  centre level critical resources like oxygen and ventilator are allocated to different states, and those items are further distributed to district authorities to be distributed to the hospital.  Clearly, the allocation of critical resources is performed at different levels, and the allocation factor is not the same at each level. The allocation should generally depend on the parameters related to active cases,   total cases,  test positive ratio, and existing resources. At the initial stages of the pandemic, the decision for the lockdown was taken only at the top level; however, at later stages, the local authorities also took the drastic decision of lockdown. The primary factors behind the resource allocation and lockdown decision are the prediction of total cases and active cases. Currently,  different physical models are adopted by different authorities; however, there is no standardized model approved by the government authority.  Also, modeling the dynamics of the pandemic is complex, and its parameter needs to be calibrated for each new area.  Therefore, sometimes the factors behind the allocation are not justifiable in a quantitative sense.   

\subsection{Decision Support System for COVID-19}
COVID-19 management at the government level is a complex task to ensure proper estimation of demand,  optimal allocation of resources, and optimal lockdown management.  The critical decision needs to be taken considering economic, medical, and complex social factors. Also, the decision-making criteria need to be formulated using quantitative tools rather than qualitative factors for fair and optimal allocation. 
As decision-making has to be performed at each level, it might not be possible for each allocation authority to model these factors quantitatively.  Currently, a COVID management authority has no standard tool to ensure optimal allocation and lockdown. Clearly, it is not easy to always make optimal decisions by a COVID-19 management authority without access to high-level technical expertise. A DSS for taking the decision of allocation and lockdown could help the government authorities in the optimal decision and avoid catastrophic failure.

In general, a decision support software should be able to integrate the information from the multiple layers of spatial and statistical databases and aid in decision making.
In the case of COVID-19 management, DSS should be designed such that it could take the decision of estimation, allocation, and lockdown management from the observed cases of total cases recovered cases and deceased cases without the requirement of complex algorithm parameters. Apart from allocation and lockdown analysis, the prediction of COVID-19  should be available to authorities for decision-making.  The prediction module should be integrated into suitable external information sources for the early prediction of disaster.

\FloatBarrier

\section{ Algorithms}  \label{algorithm}
In this section, the overview of algorithms for decision-making is discussed. Algorithms need to be designed, keeping in mind that they should be scalable and flexible. The scalability is required to ensure that DSS is applicable at lower levels consisting of the population in millions to higher-level consist of the population in tens of millions. Flexibility is required so that the algorithms are mostly independent of the region.  The three important algorithms for prediction,  allocation, and lockdown management is discussed. 
   \subsection {COVID-19 prediction}
      The various COVID-19 prediction models are available in literature based on physical modeling, data-driven modeling, and a hybrid approach that combines physical and data-driven modeling. The physical and hybrid models will require transmission parameters for each region; however, the data-driven model will need only the observed values of total cases, recovered cases, and deceased cases. Physical modeling tries to model the actual dynamics of the pandemic, but most of the pandemic-related government decisions are taken based on the reported cases.  Public perception of COVID management is mostly based on the reported cases rather than the actual COVID cases.  The data-driven model will satisfy the flexible criteria and also be simple for lower-level authorities.    We have considered a data-driven adaptive short-term model reported by \cite{jana2020adaptive} for the development of DSS.  In this case, case prediction is developed using time series data of previous observations. The prediction function is adaptively updated based on weighted least square functions to track the current dynamics of COVID-19.    The total cases, active cases, deceased cases, and active cases could be predicted using the previous observations, and the prediction is found to be reasonable up to 2-3 weeks.   As the function is adaptive and the decision frequency of administration will not be generally higher than two weeks, this algorithm could be integrated into DSS.

  \subsection{Allocation}
     Disaster management authorities allocate the lower units based on resources available from the upper administrative level and the demand requirement for the lower administrative level. The resource allocation module should be able to allocate optimally among the lower level using their demand and severity \citep{jana2021decision}. The allocation mechanism in COVID-19 is complex as the unit demand of each critical item might vary from region to region based on region-specific medical and social factors. For example, the amount of oxygen to be provided to the patient could differ depending on the medical protocol developed by the individual region.  So, the demand for medical items of low-level units is difficult to compare because of variation in protocol and unavailability of data of existing resources. So, for this, algorithms need to be selected which can incorporate those uncertain factors.  In this case, we have considered an optimization model which considers the demand of lower units and demand based on their active cases (\cite{jana2021critical}).  For allocation of each item,  the inputs are the demand of each item and the maximum value of the active cases over the next seven days. This prediction of active cases is performed using the prediction algorithm described earlier.   This algorithm provides a closed-form solution, and it is scalable for any number of units. 

\subsection{ Lockdown management}  
        
      The decision for lockdown allows less medical load, but it has a detrimental effect on the economy and various social factors.  However, it is difficult to incorporate all these factors accurately in the decision-making process as modeling of this factor is not possible at each level. To develop a simple DSS, it should be independent of those factors.    Since ensuring the availability of the medical items is the main criteria for lockdown, we have adopted a lockdown algorithm for DSS, proposed by  \cite{jana2021optimal} to ensure that there is no scarcity of medical items.  In this algorithm, the demand of each item over the next 14 days is checked, and lockdown is recommended if the availability of any of the critical items is lower than the demand at any point. The demand for the next 14 days of each item is calculated as a function of predicted active cases.

\section{ Graphical user interface design }  \label{GUI}
      
\begin{figure*}[htb!]
      \centering
      \includegraphics[width=0.45\linewidth, keepaspectratio]{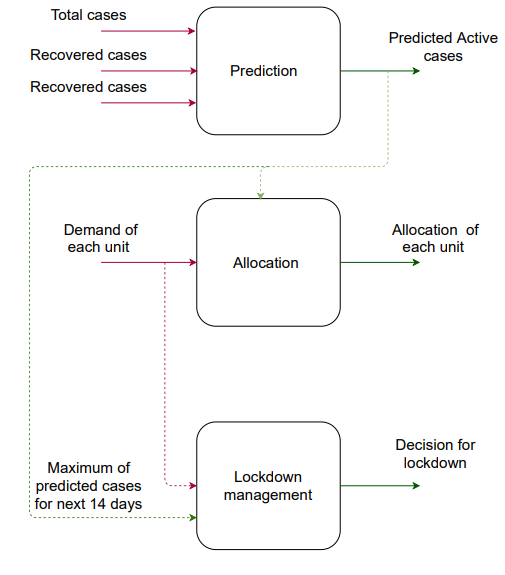}
      \caption{Input-output block diagram of GUI}
      \label{fig:COVID_DSS.png}
  \end{figure*}

Graphical user interface to be designed   so that it is easily implementable at lowest lower level of authorities with limited access to technical resources. Tentative input and output of the DSS is shown in Fig. \ref{fig:COVID_DSS.png}. Clearly, user needs to enter the COVID statistic and the demand of  critical items  for decision making process and no  specific parameter related to a particular  area is needed.

\begin{figure*}[htb!]
      \centering
      \includegraphics[width=0.8\linewidth, keepaspectratio]{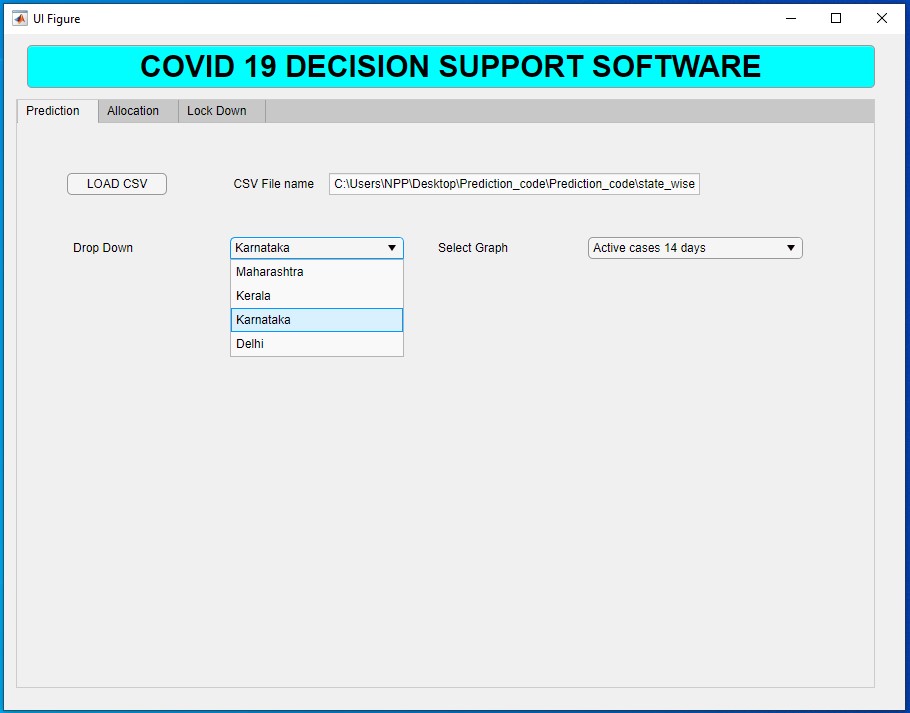}
      \caption{Prediction Tab (a)}
      \label{fig:GUI1.PNG}
  \end{figure*}

\begin{figure*}[htb!]
      \centering
      \includegraphics[width=0.8\linewidth, keepaspectratio]{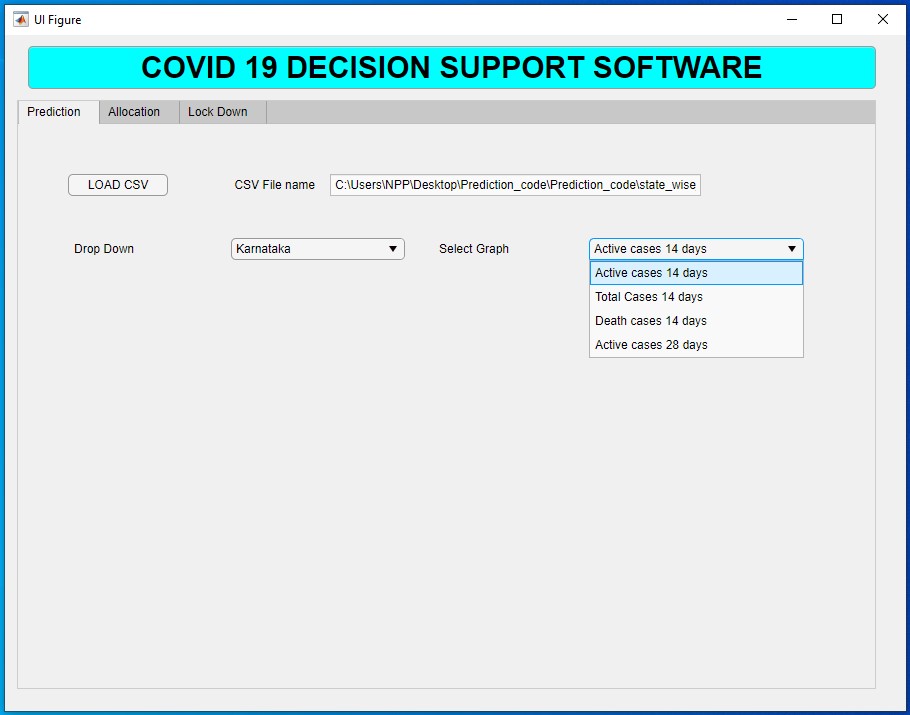}
      \caption{Prediction Tab (b)}
      \label{fig:GUI2.PNG}
  \end{figure*}

\begin{figure*}[htb!]
      \centering
      \includegraphics[width=0.8\linewidth, keepaspectratio]{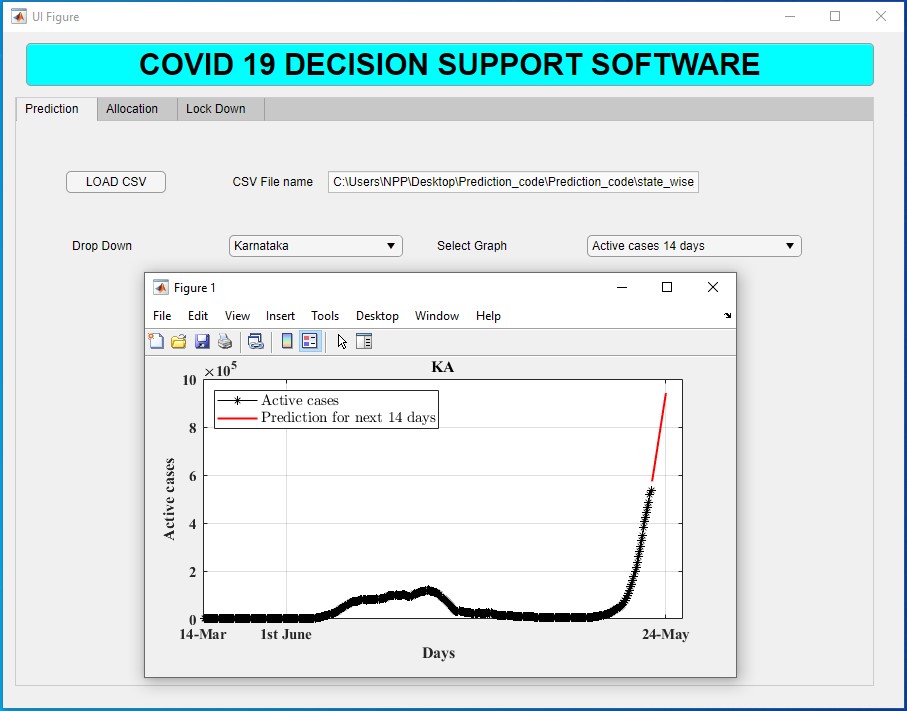}
      \caption{Prediction Tab (c)}
      \label{fig:GUI3.PNG}
  \end{figure*}

\begin{figure*}[htb!]
      \centering
      \includegraphics[width=0.8\linewidth, keepaspectratio]{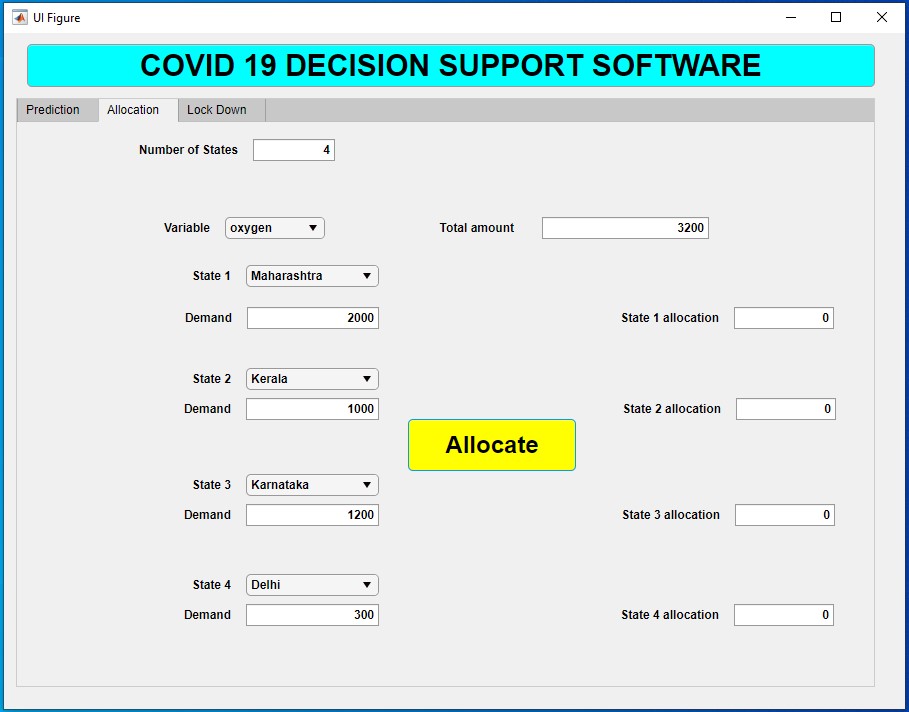}
      \caption{Allocation Tab (a)}
      \label{fig:GUI4.PNG}
  \end{figure*}

\begin{figure*}[htb!]
      \centering
      \includegraphics[width=0.8\linewidth, keepaspectratio]{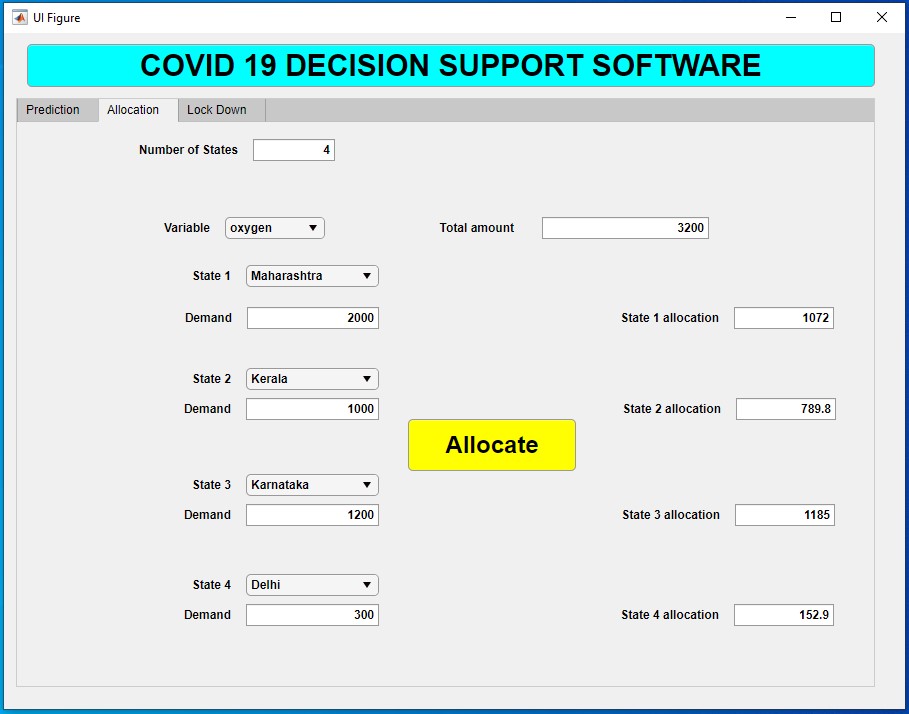}
      \caption{Allocation Tab (b)}
      \label{fig:GUI5.PNG}
  \end{figure*}

\begin{figure*}[htb!]
      \centering
      \includegraphics[width=0.8\linewidth, keepaspectratio]{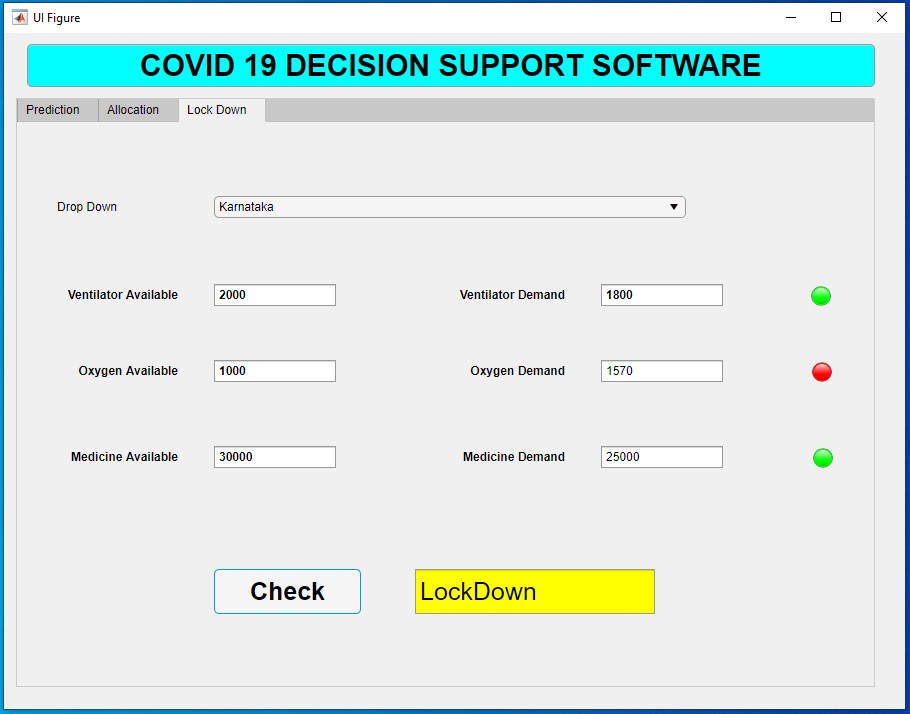}
      \caption{Lockdown Tab }
      \label{fig:GUI6.PNG}
  \end{figure*}

Generally, statistics of COVID cases are available in a specific format, and GUI has to be designed so that the output is obtained with limited manual intervention. The input data could be linked with the database of  COVID-19 to reflect the automatic change in the status. Currently,  We have developed the DSS in MATLAB  specific to the Indian context; however, it is easily adaptable for different countries. The snapshots of the different tabs of GUI are shown in Figs. \ref{fig:GUI1.PNG} -\ref{fig:GUI6.PNG}. GUI has three tabs: prediction, allocation, and lockdown. GUI is currently developed for the central government of India, and few states are considered for demonstration of GUI.  The prediction tab selects cases history from the CSV  file, and the prediction graph is generated. In the case of India, this file is directly available from the ``https://api.covid19india.org/".   Based on the states and type of graph tab,  the predicted graph is generated. The snapshots related to the prediction tab are shown in Fig. \ref{fig:GUI1.PNG} - Fig. \ref{fig:GUI3.PNG}. The GUI is currently run using the data up to 10 May 2021, and the graph in  Fig. \ref{fig:GUI3.PNG} shows the prediction of active cases of Karnataka for the next 14 days.

In the allocation tab (Fig. \ref{fig:GUI4.PNG}), the user can select the items to be allocated and the corresponding units for allocation. In this case, we have demonstrated for four states.  The user only needs to enter the corresponding demand of each state and the total amount.  The back-end algorithm predicts the active cases and incorporates that in the allocation algorithm. An allocation scenario is considered to demonstrate the effective integration of the allocation algorithm with GUI. As shown in Fig. \ref{fig:GUI4.PNG}, the allocation of oxygen is performed considering four states, and their individual demands are 2000, 1000, 1200, 300 MT. The total available oxygen for allocation is 3200 MT.  The final allocation is shown in  Fig. \ref{fig:GUI5.PNG}. The allocation algorithm considered the prediction of active cases of states on the back-end side. The final allocation is shown to be 1072, 789.8, 1185, 152.9 MT.   In the lockdown tab (Fig. \ref{fig:GUI5.PNG}) user only needs to enter the available values of the critical items.  The back-end checks using predicted demand and recommend the lockdown requirement.

   \FloatBarrier

\section{Conclusions}
 In this paper,  a decision support system to aid in the decision-making of authorities for the management of COVID-19 is presented.   The proposed  DSS consists of the prediction of COVID-19, optimal allocation, and lockdown management.  The backend algorithms for the DSS are developed based on a data-driven approach so that the proposed DSS  is applicable for any region. A MATLAB  GUI is developed incorporating the proposed DSS. Authorities could use the developed DSS for managing the allocation and lockdown-related tasks related to COVID-19.

 \FloatBarrier

\bibliographystyle{apalike}
\bibliography{main}

\end{document}